\documentclass[12pt,a4paper]{iopart}
\usepackage[dvips]{graphicx}
\expandafter\let\csname equation*\endcsname\relax
\expandafter\let\csname endequation*\endcsname\relax
\usepackage{amsmath}
\usepackage{amssymb}
\usepackage{iopams}
\usepackage{here}
\usepackage{color}
\usepackage{epsfig}
\usepackage{delarray}
\usepackage{soul}
\setstcolor{red}

\newcommand{\be}{\begin{displaymath}}
\newcommand{\ee}{\end{displaymath}}
\newcommand{\bn}{\begin{equation}}
\newcommand{\en}{\end{equation}}

\newcommand{\gyro}{{\sc gyro }}

\newcommand{\vpa}{v_{\|}}

\newcommand{\vpe}{v_{\perp}}
\newcommand{\ord}{\mathcal{O}}

\newcommand{\energy}{\mathcal{E}}

\newcommand{\Ev}{\mathbf{E}}
\newcommand{\Bv}{\mathbf{B}}
\newcommand{\mN}{\mathcal{N}}

\begin{document}
\title[Turbulent transport of impurities]{Turbulent transport of
  impurities and their effect on energy confinement}
\author{I. Pusztai${}^{1,3}$, A. Moll\'{e}n${}^1$,
  T. F\"{u}l\"{o}p${}^1$, J. Candy${}^2$ }
\address{$ˆ1$ Department of Applied Physics, Chalmers University of
  Technology and Euratom-VR Association, G\"oteborg SE-41296, Sweden}
\address{$ˆ2$ General Atomics, P.O. Box 85608, San Diego, CA
  92186-5608, USA} 
\address{$ˆ3$ Plasma Science and Fusion Center,
  Massachusetts Institute of Technology, Cambridge, MA 02139, USA}
\ead{pusztai@chalmers.se}

\begin{abstract}
By presenting linear and nonlinear gyrokinetic studies, based on a
balanced neutral beam injection deuterium discharge from the DIII-D
tokamak, we demonstrate that impurities alter the scaling of the
transport on the charge and mass of the main species, and even more
importantly, they can dramatically change the energy transport even in
relatively small quantities.  A poloidally varying equilibrium
electrostatic potential can lead to a strong reduction or sign change
of the impurity peaking factor due to the combined effect of the
in-out impurity density asymmetry and the $\mathbf{E\times B}$ drift
of impurities.  We present an approximate expression for the impurity
peaking factor and demonstrate that impurity peaking is not
significantly affected by impurity self-collisions.
\end{abstract}

\section{Introduction}
Several models have been proposed to explain the favorable isotope
mass scaling of the energy confinement in tokamaks
\cite{r5,r6,r7,r8,r9}, yet a commonly accepted and robust theoretical
explanation of the isotope effect which is consistent with the other
experimentally established parameter scalings \cite{r1,r2,r3,r4} is
still lacking. When the turbulent energy transport is determined by
the properties of the main ion species, dimensional considerations
suggest that the charge and mass scaling of the heat diffusivities,
$\chi$, should follow the gyro-Bohm scaling,
$\chi_{gB}\sim\rho_{i}^2v_{i}/a\propto \sqrt{m_i}/Z_i^2$, where
$v_i=(2T_i/m_i)^{1/2}$ is the ion thermal speed, $a$ is the plasma
minor radius, $\rho_i=v_i/\Omega_{ci}$ is the ion thermal Larmor
radius, with the cyclotron frequency $\Omega_{ci}=Z_ieB/m_i$ of an ion
of charge number $Z_i$ and mass $m_i$, and $B$ is the magnetic field
strength.  Studies of charge and mass effects on the turbulent energy
transport in DIII-D tokamak plasmas show that the presence of
impurities can contribute to deviations from a pure gyro-Bohm scaling
\cite{pusztaicandy}.  In this paper we use the expression ``pure
gyro-Bohm'' if differences in the turbulence between similar plasmas
with different main species can be exactly transformed out by
normalizing temporal and spatial scales to $a/v_i$ and $\rho_i$,
respectively. In the first part of the paper we present
\emph{similarity studies}, that is, a series of gyrokinetic (GK)
simulations with {\sc gyro}~\cite{gyro} where the local geometry and
profile parameters -- taken from a deuterium discharge studied in
Ref.~\cite{pusztaicandy} -- are kept fixed while the ion composition
is artificially changed from case to case.  First the effect of
impurities (in particular, carbon or deuterium) on energy transport in
hydrogen isotope and helium plasmas is studied.  Then we investigate
the role of the main species on impurity transport, and conclude that
impurity transport does not follow the pure gyro-Bohm scaling (based
on the properties of the main ion species) so closely as the energy
transport.

One of the ways to reduce impurity accumulation in tokamak cores is to
apply central radio frequency (RF) heating
\cite{valisa,duxppcf,puiatti06,neu05}. RF-heating can generate
poloidal impurity density asymmetries \cite{kazakov} and if the
asymmetry is sufficiently strong, it can lead to a sign change in the
impurity peaking factor~\cite{poloidal,poloidal2}. Recently it was
shown that a sign change of the impurity peaking factor can happen at
much weaker (that is, realistic) asymmetry strengths than were
considered in Ref.~\cite{poloidal2} if the $\Ev_\theta\times \Bv_\phi$
drift of impurities in the poloidally varying equilibrium
electrostatic field is taken into account~\cite{albert}.  In the
second part of this paper, we extend the work of Ref.~\cite{albert},
by including parallel dynamics and finite Larmor radius (FLR) effects
and modeling impurity self-collisions with the full linearized
operator for impurity self-collisions to derive an improved
approximate expression for the impurity peaking factor.

\section{Charge and mass effects on turbulent transport}
\label{sec:chargemass}
In this section we study how the charge and mass scalings of the
energy transport are affected by the presence of impurities, through
linear and nonlinear GK similarity studies with {\sc gyro}, based on
local profile and magnetic geometry data from an L-mode phase of a
weakly rotating DIII-D deuterium discharge (129135, 1250-1300 ms).
For detailed information on the resolution of the simulations and on
the studied discharge we refer the reader to
Ref.~\cite{pusztaicandy}. The dominant impurity in the discharges
studied in Ref.~\cite{pusztaicandy} was carbon, and the hydrogen and
helium discharges had a significant deuterium fraction present; this
motivates our choice of impurities in the simulations presented in
this section. We neglect particle flows within the flux surface and to
model different ion compositions we keep the electron density profile
fixed and assume constant ion concentrations. For different main
species with the same charge the results of local GK simulations --
normalized to main ion species units -- should exactly coincide when
impurities and collisions are neglected and adiabatic electron
response is assumed~\cite{pusztaicandy}. Thus we normalize growth
rates, $\gamma$ to $c_{si}/a$, where $c_{si}=(T_e/m_i)^{1/2}$,
binormal wave numbers $k_y$ to $1/\rho_{si}=\Omega_{ci}/c_{si}$, and
fluxes to gyro-Bohm units $Q_{gBi}=n_eT_ec_{si}(\rho_{si}/a)^2$,
$\Gamma_{gBi}=n_ec_{si}(\rho_{si}/a)^2\propto m_i^{1/2}/Z_i^2$.

\begin{figure}[ht]
\begin{center}
\includegraphics[scale=1.05]{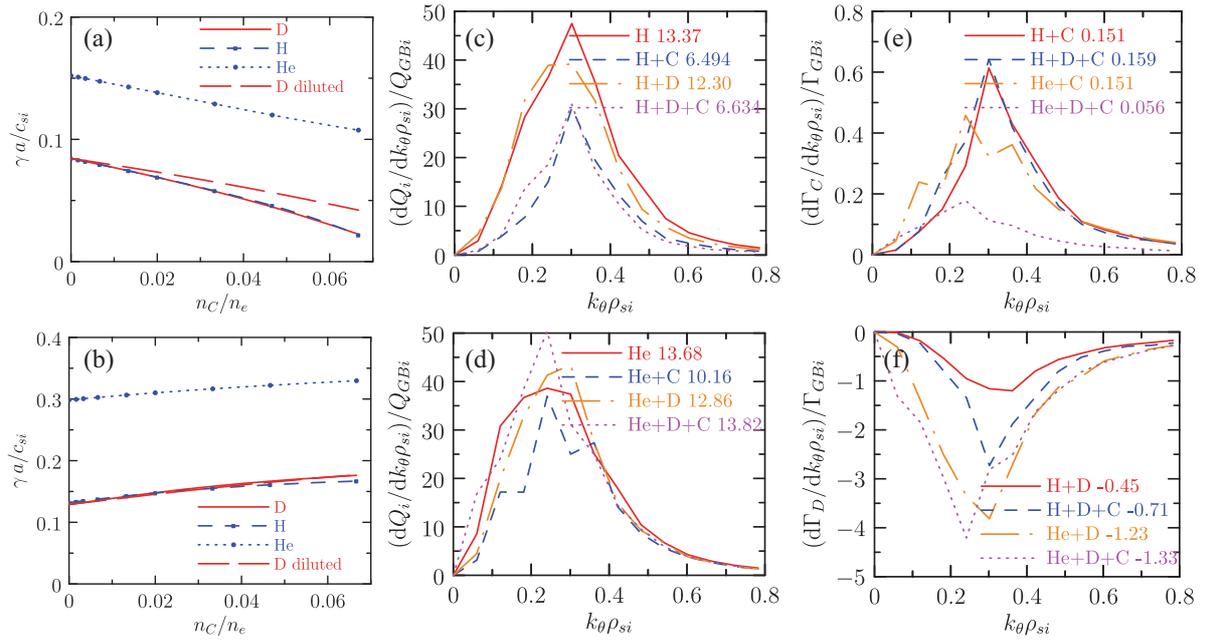}
\caption{(a,b) Growth rates of the $k_y\rho_{si}=0.3$ mode as a
  function of carbon concentration. Solid: main ion species is
  deuterium, dashed: hydrogen, dotted: helium, long dashed: diluted
  deuterium (no carbon, non-quasineutral). (c,d) Ion energy flux
  spectra [in D units] of pure plasmas (solid), with carbon (dashed),
  with deuterium (dash-dotted), and with carbon and deuterium
  impurities. (c): main ion hydrogen, (d): helium. (e) and (f): carbon
  and deuterium particle flux spectra. }
 \label{fig:simil}
\end{center}
\end{figure}

In Figs.~\ref{fig:simil} (a,b) the growth rate of linear modes at
$k_y\rho_{si}=0.3$ (at $r/a=0.55$) is shown as a function of carbon
concentration $n_C/n_e$. The local geometry and profile parameters, in
the notation of Ref.~\cite{gyro}, are: $q=1.78$, $s=0.75$,
$\kappa=1.34$, $\delta=0.11$, $a/L_n=0.68$, $a/L_{Ti}=1.89$,
$a/L_{Te}=2.69$, $T_i/T_e=0.82$, $\nu_{ei}=0.14\;c_s/a$.  The solid
line corresponds to a deuterium plasma, the dashed is a hydrogen and
the dotted is a helium plasma. We also show simulations for diluted
deuterium plasma (long dashed lines), i.e., the deuterium
concentration is decreased to the value that would correspond to the
given carbon concentration (according to $n_i/n_e=1-Z_C n_C/n_e$), but
the carbon has no response to fluctuations (the carbon density is set
to zero in the simulation). Fig.~\ref{fig:simil} (a) shows the
simulation results with adiabatic electrons.  The decrease in the
growth rates with increasing carbon concentration is practically
linear. For lower carbon concentrations the slope of the $\gamma$
curves is the same in main ion species units, independently of the
mass or charge of the main species. For higher carbon concentrations a
small difference between hydrogen isotopes and helium appears
regarding the slopes, but the deuterium and hydrogen curves still
overlap almost exactly, in spite of the difference between the
$m_C/m_i$ mass ratios. Noticeable deviations between growth rates in
different hydrogen isotope plasmas appear only at higher wave numbers;
noting that $1/\rho_{sC}\approx 2.45/\rho_{si}$.

Figure~\ref{fig:simil} (b) shows trapped electron (TE) mode growth
rates in a collisionless drift-kinetic electron simulation. The TE
modes are weakly destabilized as the impurity concentration is
increased and this effect is somewhat stronger in hydrogenic
plasmas. This destabilization is due to the increased weight of
trapped-electrons in the Poisson equation due to the diluting effect
of impurities (similarly, the stabilizing effect of impurities on ITG
modes mentioned before is mostly due to the decreased weight of
ions). But it has to be noted that these TE modes are strongly
stabilized when electron-ion collisions are introduced in the
simulation, and the stabilizing effect of increasing $Z_{\rm eff}$
appearing in the electron-ion collision frequency can be much stronger
than the destabilizing effect observed in the collisionless case.

Finally, we compare the deuterium plasma simulations with the correct
physics (solid lines) to the model where only the diluting effect of
impurities is taken into account (long dashed lines). The presence of
carbon contributes to a further stabilization of the ITG modes, and
this contribution is comparable to the effect of the dilution of the
main species. However, for the TE mode, the presence of carbon has
almost negligible effect compared to that caused by the dilution of
the main species.

Moving on to nonlinear simulations, the curves in
Figs.~\ref{fig:simil}~(c-f) represent the distribution of energy (or
particle) fluxes over poloidal wave numbers; the $k_\theta\rho_{si}$
integral of the curves give the fluxes in gyro-Bohm units; the
corresponding flux values are given in the plot legends. The nonlinear
simulations include kinetic electron response and electron-ion
collisions and they are based on the local parameters of the studied
deuterium plasma at $r/a=0.65$ ($q=2.08$, $s=1.1$, $\kappa=1.36$,
$\delta=0.14$, $a/L_n=0.79$, $a/L_{Ti}=1.94$, $a/L_{Te}=3.07$,
$T_i/T_e=0.91$, $\nu_{ei}=0.23\;c_s/a$).

First, we study the respective and combined effects of the carbon and
deuterium impurities on energy transport. Figure~\ref{fig:simil} (c)
shows the \emph{total} ion energy flux in species units for a pure
hydrogen plasma (solid curve), in hydrogen plasmas with carbon
(dashed) and deuterium (dash-dotted) impurities, and with both carbon
and deuterium impurities (dotted); $n_C/n_i=0.04167$ corresponding to
$Z_{\rm eff}=2$ in hydrogenic plasmas, and $n_D/n_i=1/3$. The
stabilizing effect of carbon on ITG modes, seen also in the linear
simulations, appears as a strong reduction (by $51 \%$) of the ion
energy fluxes that acts at all wave numbers (we note that TE modes are
stabilized by collisions). The presence of deuterium slightly reduces
the transport at higher wave numbers (this feature is consistent with
corresponding linear simulation results, not shown
here). Interestingly, when there is carbon impurity in the plasma, the
presence of a deuterium minority does not reduce the absolute value of
the transport, only the spectrum gets shifted slightly towards lower
wave numbers. The effect of these impurities is qualitatively the same
on the electron energy flux, the difference is that the reduction in
the electron energy transport is not as strong as in the ion energy
transport (only $35\%$). Figure~\ref{fig:simil} (d) shows how the
presence of D and C affects the energy transport in a helium
plasma. The effect of a C impurity is considerably weaker in a helium
plasma than in hydrogen isotope plasmas and it is concentrated to
lower wave numbers. The fact that we kept $n_C/n_i$ fixed in the
simulations [shown in Figs.~\ref{fig:simil} (c) and (d)] instead of
$n_C/n_e$ is part of the reason why we find a weaker effect of carbon
in the simulation with helium than with hydrogen main ions, since then
carbon density is smaller in the helium plasma. However, even if we
kept $n_C/n_e$ fixed, looking at Fig.~\ref{fig:simil} (a) we expect to
have a weaker impact of the carbon impurity on transport because the
relative change in the linear growth rates (compared to the pure
plasma case) for given carbon concentration is smaller in a helium
plasma than in hydrogen. While both carbon and deuterium act to reduce
the energy transport in the helium plasma (compare the gyro-Bohm
normalized fluxes in the legend of Fig.~\ref{fig:simil} (d)),
interestingly, when both impurities are present the absolute value of
the transport remains almost the same as in the pure plasma, however
the shape of the energy flux spectra changes somewhat, being more
peaked around its maximum.

In Figs.~\ref{fig:simil} (e,f) the carbon and deuterium particle
fluxes are shown in impure hydrogen and helium plasmas, respectively
(positive sign corresponding to an outward particle flux).  In species
units the magnitude of the carbon particle flux is the same in a
hydrogen and a helium plasma when the ratio $n_C/n_i$ is kept fixed;
compare the solid and dash dotted curves in Fig.~\ref{fig:simil}
(e). The shape of the carbon particle flux spectra are similar to the
ion energy flux spectra in the hydrogen and helium plasmas [H+C and
  He+C curves in Figs. (c,d) compared to those in Fig. (e)], however,
the magnitude of $Q_i$ is smaller in the H+C than in the He+C plasma.
It is also interesting to note, that in the hydrogen plasma the carbon
particle flux remains almost unaffected when a deuterium minority is
introduced; in contrast to the helium plasma, where the carbon flux
drops dramatically (by $63\%$) in the presence of deuterium; a trend
opposite to what was found for ion energy fluxes.

Figure~\ref{fig:simil} (f) shows that the deuterium particle fluxes in
species units appear to be almost three times higher in a helium
plasma than in a hydrogen plasma, even if the $n_D/n_i$ values are
held constant, so the number of deuterium particles is half as many in
the helium plasma as in the hydrogen plasma. In spite of the fact that
the linear growth rates and the energy transport are reduced when
carbon impurity is present, the deuterium particle fluxes increase
both in a H(+D) and a He(+D) plasma when carbon impurity is
introduced.

\section{Impurity peaking factor in the presence of poloidal asymmetries}
\label{sec:asym}
In this section we calculate the zero flux density gradient (peaking
factor) of trace impurities under the effect of a poloidally varying
equilibrium electrostatic potential $\phi_E$ that can arise due to
e.g. the presence of a species with strong temperature anisotropy
\cite{reinke}. Especially, in case of RF heating on the outboard side,
the trapped population of the heated particle species increases in the
vicinity of the resonance position \cite{ingessonppcf,kazakov}. The
peaking factor depends on the linear mode characteristics and on the
form of $\phi_E$. Here, we neglect effects caused by a radial electric
field, such as toroidal rotation.

We assume $e\Delta \phi_E/T_j \ll 1$, where $\Delta \phi_E$ refers to
the poloidal variation of $\phi_E$, so that poloidal asymmetry effects
on the main species can be neglected; this assumption is needed to
justify the use of {\sc gyro} simulations to obtain the linear mode
characteristics.  However, we allow for $Z e \Delta
\phi_E/T_z=\ord(1)$.

\subsection{Perturbed impurity distribution}
\label{sec:distrib}
We consider the particle transport driven by a single, representative,
toroidal mode. The impurity peaking factor is calculated by requiring
the linear impurity flux $\Gamma_z\equiv\Im\left\langle
-k_y\hat{n}_z\phi^\ast/B\right\rangle = \Im\left\langle-k_y\int d^3v
J_0(z_z) g_z \phi^\ast/B\right\rangle $ to vanish (steady state is
assumed and impurity sources are neglected). Here $\langle\cdot
\rangle $ denotes a flux surface average, $\hat{n}_z$ is the perturbed
impurity density, $g_z$ the non-adiabatic part of the perturbed
impurity distribution function, $J_0$ is the Bessel function of the
first kind, $z_z=k_\perp v_{\perp}/\omega_{cz}$, $k_\perp = \left(1 +
s^2 \vartheta^2\right)^{1/2} k_y$, and the rest of the notation is
standard.  The species indexes $z$, $e$ and $i$ refer to impurities,
electrons and main ions.

The linearized GK equation is to be solved to obtain the non-adiabatic
part of the impurity distribution function $g_z$,
\begin{equation}
  \left.\frac{v_\parallel}{q R} \frac{\partial {g}_z}{\partial
    \vartheta}\right|_{\mathcal{E},\mu}-i(\omega-
  \omega_{Dz}-\omega_E) {g}_z - C[g_z] =-i\frac{Z e
    f_{z0}}{T_z}\left(\omega-\omega_{\ast z}^T \right)\phi J_0(z_z),
\label{gke}
\end{equation}
where $\vartheta$ is the extended poloidal angle, $\omega = \omega_r +
i \gamma$ is the mode frequency,
$f_{z0}(\psi,\energy)=n_{z0}(\psi)[m_z/2\pi
  T_z(\psi)]^{3/2}\exp[-\energy/T_z(\psi)]$ is the equilibrium
Maxwellian distribution. $\mu = m_z \vpe^2/\left(2 B\right)$, $\energy
= m_z v^2/2 + Z e \phi_E$, $\omega_ {\ast z}^T=\omega_{\ast z}\left[1
  - L_{nz} Z e \partial_r( \phi_E
  e\phi_E/T_z-3/2)L_{nz}/L_{Tz}\right] $ with $\omega_{\ast z}=-k_y
T_z/Z eB L_{nz}$, $L_{nz}=-\left[\partial_r\left(\ln{n_z}\right)
  \right]^{-1}$ and $L_{Tz}=-\left[\partial_r\left(\ln{T_z}\right)
  \right]^{-1}$.  Furthermore, $ \omega_{Dz}=-2 k_y /m_z(\energy-Z
e\phi_E-\mu B/2) \mathcal D(\vartheta)/\omega_{cz} R$, where $\mathcal
D(\vartheta)= \cos{\vartheta}+ s \vartheta \sin{\vartheta}$. The
$\Ev\times \Bv$ drift frequency in the equilibrium electrostatic field
is $\omega_{E}=k_y \left[\partial_r \phi_E-s \vartheta
  \partial_\vartheta( \phi_E)/r\right]/B$.  Our ordering
$Ze\phi_E/T_z\sim\ord (1)$ requires that $\omega_E/\omega$ is formally
$\sim\!1/Z$ small; this requirement is mostly satisfied for
experimentally relevant poloidal asymmetries. We consider ion scale
modes, $z_i\lesssim 1$, and we allow the parallel streaming term of
main ions to be comparable to the mode frequency.  In the following we
perform a perturbative solution of the impurity GK equation
(\ref{gke}) in the small parameter $Z^{-1/2} \ll 1$. We assume that
impurity self-collisions dominate over unlike-particle collisions,
requiring $n_zZ^2/n_e=\ord(1)$, thus we model only impurity-impurity
collisions with the full linearized collision operator
$C^{(l)}_{zz}[\cdot]$.

 Keeping in mind that $\omega_E/\omega$, $\omega_{Dz}/\omega$,
 $\omega_{\ast z}^T/\omega$, and $J_0(z_z)-1\approx -z_z^2/4$ are all
 $\sim\!1/Z$ small, the lowest order equation $ -i\omega g_0 -
 C^{(l)}_{zz}[g_0]=-i\omega Z e \phi f_{z0}/T_z $ for
 $g_z=g_0+g_1+g_2+\dots$ is satisfied by $g_0=Z e \phi f_{z0}/T_z $,
 since $C^{(l)}_{zz}[g_0\propto f_{z0}]$ vanishes.  To first order in
 $Z^{-1/2}$, we have $v_\parallel \partial_\vartheta(g_0)/(q R) -
 i\omega g_1 - C^{(l)}_{zz}[g_1]=0$. Using $C^{(l)}_{zz}[g_1\propto
   \vpa f_{z0}]=0$, and $\partial_\theta(f_{z0})|_\energy=0$, we
 obtain $ g_1=-i Ze f_{z0} v_\parallel \partial_\vartheta
 (\phi)/(T_z\omega q R)$.  To next order, the GK equation reads
\begin{equation}
- i\omega g_2 - C^{(l)}_{zz}[g_2]=-i
(\omega_{Dz}+\omega_E)g_0-v_\parallel \partial_\vartheta
({g}_1)/(qR)+i Z e \phi f_{z0}(\omega z_z^2/4+\omega_{\ast z}^T )/T_z.
\label{gk2s}
\end{equation}
Using that $m_z\vpa \partial_\vartheta(\vpa)|_{\energy,\mu}=-\mu
\partial_\vartheta B-Z e \partial_\vartheta \phi_E$, the parallel
compressibility term in the right hand side of Eq.~(\ref{gk2s}), can
be expressed as
\begin{equation}
-\frac{\vpa}{qR}\frac{\partial g_1}{\partial \vartheta}= i \frac{ Z e
  f_{z0} }{\omega T_zq^2R}\left\{\vpa^2
\frac{\partial}{\partial\vartheta}\left( \frac{1}{R}\frac{\partial
  \phi}{\partial\vartheta}\right) -\frac{1}{R} \frac{\partial
  \phi}{\partial\vartheta} \left[\frac{v_\perp^2}{2}\frac{\partial \ln
    B}{\partial\vartheta} + \frac{Ze}{m_z}\frac{\partial
    \phi_E}{\partial\vartheta}\right]\right\}.
\label{parcterm}
\end{equation}
We decompose the magnetic drift frequency and the FLR parameter as
$\omega_{Dz}=\omega_{Dx}+\omega_{D\|}$ and $z_z^2=z_x^2-z_\|^2$, where
$\omega_{Dx}$ and $z_x^2$ are proportional to $v^2$ and $\omega_{D\|}$
and $z_\|^2$ are proportional to $v_\|^2$. Also, we write
$g_2=\hat{g}_2+\tilde{g}_2$, where $\hat{g}_2$ has contributions
proportional to $f_{z0}$ and $v^2 f_{z0}$, thus
$C^{(l)}_{zz}[\hat{g}_2]=0$.  The collisionless, isotropic part of
Eq.~(\ref{gk2s}) is solved by
\begin{equation}
\hat{g}_2= \frac{\omega_{Dx}+\omega_E}{\omega}g_0 - \frac{Z e
  \phi}{T_z}f_{z0}\frac{\omega z_x^2/4+\omega_{\ast
    z}^T}{\omega}+\frac{Ze f_{z0}}{\omega^2 T_z
  q^2R^2}\frac{\partial\phi}{\partial
  \vartheta}\left[\frac{v^2}{2}\frac{\partial \ln
    B}{\partial\vartheta} +\frac{Ze}{m_z}\frac{\partial
    \phi_E}{\partial\vartheta}\right].
\label{g2hatss1}
\end{equation}
Then $\tilde{g}_2$ should satisfy the remaining part of
Eq.~(\ref{gk2s}) which, upon division by $-i\omega$ reads
\begin{equation}
\tilde{g}_2 -\frac{i}{\omega}
C^{(l)}_{zz}[\tilde{g}_2]=\frac{\omega_{D\|}}{\omega}g_0+\frac{Z e
  \phi}{T_z}f_{z0}\frac{z_\|^2}{4}-\frac{Ze v_\parallel^2
  f_{z0}}{\omega^2 T_z q^2 R} \left\{\frac{\partial }{\partial
  \vartheta} \left[\frac{1}{R} \frac{\partial \phi}{\partial
    \vartheta}\right]+\frac{1}{2R}\frac{\partial \phi}{\partial
  \vartheta}\frac{\partial \ln B}{\partial \vartheta}\right\}.
\label{gk2tildes}
\end{equation}
Equation~\ref{gk2tildes} can be written as $\tilde{g}_2
-\frac{i}{\omega} C^{(l)}_{zz}[\tilde{g}_2]-X\vpa^2f_{z0}=0$, where
$X$ is independent of velocity.  The solution can be written in the
form $\tilde{g}_2=\Xi(\mathbf{v}) f_{z0}$, where $\Xi $ can
incorporate any nontrivial velocity dependence.  Taking the density
moment of Eq.~(\ref{gk2tildes})
\begin{equation}
\int d^3v\left[ \tilde{g}_2 -i
  C^{(l)}_{zz}[\tilde{g}_2]/\omega-X\vpa^2f_{z0}\right]=0.
\label{gk2tildens}
\end{equation}
Since $ \int d^3v C^{(l)}_{zz}[\Xi(\mathbf{v}) f_{z0}]=\int d^3v \Xi
(\mathbf{v}) C^{(l)}_{zz}[f_{z0}]= 0$, Eq.~(\ref{gk2tildens}) can be
reduced to $\int d^3v\left[ \tilde{g}_2-X\vpa^2f_{z0}\right]=0$.

Since we are only interested in the density moment of the perturbed
distribution function needed to evaluate the impurity peaking factor,
instead of solving the complicated collisional
equation~(\ref{gk2tildes}) unnecessarily, we derive a substitute
function $\tilde{g}_{2s}$ that has the same density moment as
$\tilde{g}_2$; we choose $\tilde{g}_{2s}=X v^2f_{z0}/3$.

We introduce a substitute function for $g_z$, denoted by $g_s$, with
the property $\int d^3v J_0(z_z)g_s=\int d^3v J_0(z_z) g_z+
\ord(Z^{-2})$; we define it as $g_{s}=g_0+\hat{g}_2+\tilde{g}_{2s}$,
that reads
\begin{equation}
\frac{g_{s} T_z}{Z e f_{z0}} =
\phi\left[1+\frac{4\omega_{Dx}/3+\omega_E-\omega_{\ast z}^T}{\omega}
  -\frac{z_x^2}{6}\right]-\frac{v_z^2}{3 (\omega q R)^2} \left[ x_z^2
  \frac{\partial^2\phi}{\partial
    \vartheta^2}-\frac{3}{2}\frac{\partial\phi}{\partial
    \vartheta}\frac{\partial}{\partial
    \vartheta}\left(\frac{Ze\phi_E}{T_z}\right)\right],
\label{g2hatss}
\end{equation}
where $x_z=v/v_z$, with $v_z=(2T_z/m_z)^{1/2}$.  It is important to
emphasize, that $g_s$ is \emph{not} an approximate solution to
Eq.~(\ref{gke}), but merely a function that has the same density
moment as the solution. Taking other velocity moments of this function
would give erroneous results, and $g_s$ cannot be used to proceed with
the perturbative solution.

\subsection{Zero flux impurity density gradient}
\label{sec:peaking}
To calculate the impurity peaking factor we assume a simple sinusoidal
poloidal asymmetry of the form $Ze\phi_E/T_z=-\kappa
\cos(\theta-\delta)$ where $\kappa$ is the asymmetry strength and
$\delta$ determines the poloidal position of the impurity
accumulation. When evaluating the density moment of
Eq.~(\ref{g2hatss}), we keep the $\ord(Z^{-1})$ correction from
$J_0(z_z)$ where it multiplies $g_0$. Also, we neglect
$\ord(\epsilon)$ corrections when solving for $a/L_{nz}^0$, the value
of $a/L_{nz}$ where $\langle\Gamma_z\rangle =0$, to find
\begin{equation}
\frac{a}{L_{nz}^0}= 2 \frac{a}{R_0}\langle \mathcal{D} \rangle_\phi +
\frac{a}{r}s\kappa\langle\theta \sin(\theta-\delta) \rangle_\phi -
\frac{2 a v_i}{(q R_0)^2k_y\rho_i}\frac{Z
  m_i}{m_z}\frac{\omega_r}{\omega_r^2+\gamma^2} \left\langle \left|
\frac{\partial \phi}{\partial \theta}\right|^2/
\left|\phi\right|^2\right\rangle_\phi,
\label{peak1}
\end{equation} 
where $\rho_i=v_im_i/eB_0$, and $\langle\dots\rangle_\phi=\langle\dots
\mN |\phi|^2\rangle/\langle\mN |\phi|^2\rangle$, with $\mN
=\exp[\kappa \cos(\theta- \delta)]$. The FLR terms do not appear in
$a/L_{nz}^0$, since their imaginary part is zero.  The first and
second terms of Eq.~(\ref{peak1}) represents the contributions from
$\omega_D$ and $\omega_E$, respectively. Since the last term of
Eq.~(\ref{peak1}) contains only non-negative quantities, except
$\omega_r$, impurity parallel dynamics acts to increase(decrease) the
impurity peaking if $\omega_r$ is negative(positive). Note, that we
use the sign convention of {\sc gyro}; $\omega_r$ is negative for
modes propagating in the ion diamagnetic direction. The impurity
parallel compressibility term introduces a dependence on mode
frequency, and also on the charge to mass ratio, consistently with
Ref.~\cite{angionip}.

\begin{figure}[th]
\begin{center}
\includegraphics[scale=0.8]{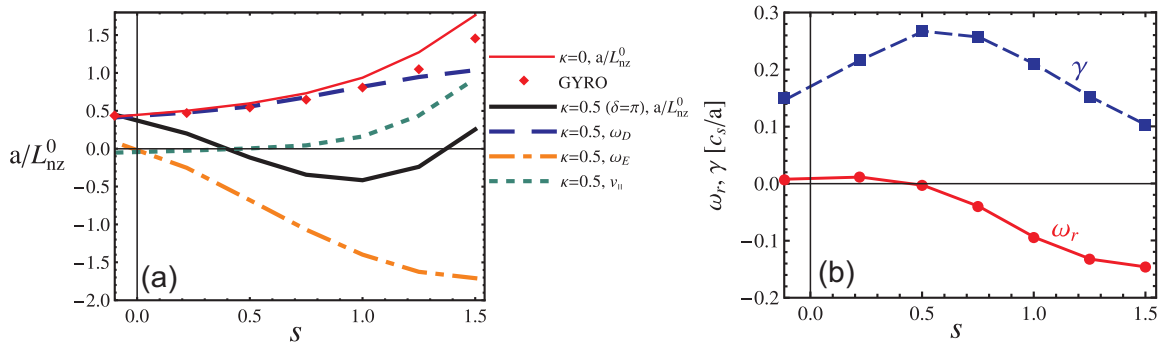}
\caption{(a) Shear dependence of the impurity peaking factor. Thin
  solid line represents the poloidally symmetric case; the
  corresponding \gyro results are shown with red diamonds.  Thick
  solid line represents an in-out asymmetry, $\kappa=0.5$. The
  contributions from the different terms in Eq.~(\ref{peak1}) are also
  plotted; dashed line: magnetic drifts, dash-dotted: $\Ev\times\Bv$
  drifts, dotted line: parallel compressibility. (b) Shear dependence
  of $\omega_r$ and $\gamma$ from \sc{gyro}.}
 \label{fig:shearscan}
\end{center}
\end{figure}

The shear dependence of the peaking factor of a fully ionized Nickel
impurity calculated from Eq.~(\ref{peak1}) is shown in
Figure~\ref{fig:shearscan}(a). The local profile and geometry
parameters assuming a circular model Grad-Shafranov equilibrium are:
$r/a=0.3$, $R_0/a=3$, $k_\theta\rho_s=0.3$, $q=1.7$, $a/L_{ne}=1.5$,
$T_i/T_e=0.85$, $a/L_{Te}=2$ and $a/L_{Ti}=2.5$, and collisions are
switched off in the simulations to avoid discrepancies due to the
different collision operators in this paper and in {\sc gyro}. The
poloidally symmetric case is shown with thin solid line and shows a
relatively good agreement with the exact values obtained from \gyro
simulations represented by the diamond markers; the differences
towards higher magnetic shear appear as a result of an overestimation
of the parallel compressibility term due to the approximations done in
the derivation of Eq.~(\ref{peak1}).  A scenario with in-out impurity
asymmetry for the asymmetry strength $\kappa=0.5$ is also plotted with
thick solid line, together with the contributions from the different
terms in Eq.~(\ref{peak1}). Around $s=0$ the contribution from
$\omega_E$ (dash-dotted line) vanishes, and in this specific case the
parallel compressibility term (dotted line) is also small due to the
small value of $\omega_r$; the peaking factor is dominated by
$\omega_D$ (dashed line). If $|\omega_r|/\gamma\ll 1$, that happens
for lower shear in the case studied here [see Fig.~\ref{fig:shearscan}
  (b)], the frequency dependent part of the third term in
Eq.~(\ref{peak1}) is small, thus parallel compressibility effects do
not play an important role; see the dashed curve in
Fig.~\ref{fig:shearscan} (a).

As the magnetic shear increases the $\omega_E$ term starts to reduce
the peaking factor, since $\langle\theta\sin(\theta-\pi)\rangle_\phi$
is negative for moderately ballooning eigenfunctions. In
Eq.~(\ref{peak1}) the shear dependent part of $\mathcal{D}$, and the
$\omega_E$ term can be combined as $(a s/R_0)\langle \theta \sin\theta
\rangle_\phi (2\pm\kappa/\epsilon)$, where $\epsilon=r/R_0$, and the
plus(minus) sign applies for out-in(in-out) impurity asymmetry. Thus,
if $\phi(\theta)$ and $\omega$ are only weakly dependent on shear,
increasing shear is expected to reduce the impurity peaking in the
in-out asymmetric case when $2<\kappa/\epsilon$ [for our simulation
  parameters $\kappa/\epsilon=5$]. From Eqs.~(10) and (12) in
Ref.~\cite{albert} one can estimate the ratio of the asymmetry
strength and the inverse aspect ratio; assuming $T_e\approx T_i\approx
T_z$, and taking the limits $\alpha_T\gg 1 $ and $\epsilon \ll 1$ so
that $\epsilon \alpha_T$ is also small we obtain
$\kappa/\epsilon\approx \alpha_T Z X_m/(1+Z_{\rm eff})$, where
$\alpha_T$ is the minority temperature anisotropy, and $X_m$ is the
minority fraction.

\section{Conclusions}
\label{sec:conclusions}

In the present paper we first study how impurity transport is affected
by the charge and mass of the main ion species and the ion
composition, and what impact the impurities have on energy transport,
then we consider the effect of a poloidally varying equilibrium
potential on impurity peaking.

Linear gyrokinetic analysis shows that for increasing impurity
concentration the growth rates change in a linear fashion; in main ion
species units the change in the growth rates when increasing impurity
concentration to a given value -- i.e. the slope of the $\gamma a
/c_{si}(n_C/n_e)$ curves -- is almost independent of the charge and
mass of the main species. In the studied case, based on local profile
and geometry data in a deuterium discharge from DIII-D, for ITG modes
carbon impurity has a strongly stabilizing,-- for TE modes (when
neglecting collisions) a weak destabilizing -- effect. For TE modes
the effect is almost purely due to plasma dilution, for ITG modes the
presence of the impurity itself has a comparable effect to the
dilution that still dominates. The strong stabilizing effect of carbon
in hydrogenic plasmas appears in nonlinear simulations as a strong
reduction of energy transport, the relative change in the energy
fluxes being even higher than that of the linear growth rates. The
effect of impurities on energy transport is smaller in a helium
plasma, due to the higher species units growth rates and the lower
main ion concentration. The presence of deuterium mainly just changes
the shape of the energy flux spectrum, although slightly reduces the
transport in a hydrogen plasma. In general the presence of impurities,
even in relatively small quantities, can cause significant deviations
from a pure gyro-Bohm mass and charge scaling of energy
fluxes. (Again, by ``pure gyro-Bohm scaling'' we refer to the
situation when differences in the turbulence between similar plasmas
with different ion composition can be exactly transformed out by
normalizing temporal and spatial scales to $a/v_i$ and $\rho_i$,
respectively. We do not mean that the transport is not gyro-Bohm in
the general sense.)

We conclude that, while the apparent deviations from a pure gyro-Bohm
scaling, -- due to charge, electron-to-ion mass ratio or collisional
effects, and importantly due to the presence of impurities, -- can
mostly be explained and understood from a linear analysis of the
underlying microinstabilities, impurity transport do not follow these
naive expectations. In particular, if we would normalize the impurity
fluxes to ion energy fluxes we would get significantly different
values in plasmas with different main ion species. Thus we might
expect that the steady state impurity profiles would also vary in
these plasmas (especially when impurity sources and neoclassical
impurity transport are accounted for), causing further deviations from
a pure gyro-Bohm scaling.

A poloidally varying electrostatic field, appearing due to e.g.~a
poloidally asymmetrically distributed RF heated particle species, even
if being too weak to modify the dynamics of the main species, can
essentially change impurity transport compared to a poloidally
symmetric situation. We show that the combined effect of the arising
poloidally asymmetric impurity distribution and the $\Ev_\theta\times
\Bv_\phi$ drift of impurities can lead to a reduction or even a sign
change in the impurity peaking factor. The effect becomes important at
high impurity charge, when the magnetic and diamagnetic drifts
(proportional to $1/Z$) become as small as the $\Ev_\theta\times
\Bv_\phi$ drift in the poloidally varying equilibrium potential. We
demonstrate that to lowest order in $1/Z$ finite Larmor radius effects
and impurity collisions do not affect impurity peaking driven by ion
scale microinstabilities as long as the impurity collisions are
dominated by self-collisions and the impurity collision frequency is
not much larger than the mode frequency. We present and analyze a
simple analytical expression for the impurity peaking factor including
contributions from the $\Ev_\theta\times \Bv_\phi$ drift and parallel
compressibility, and depending on the linear mode characteristics. We
find that to get sign change in the impurity peaking factor, a
necessary criterion is that the ratio of the asymmetry strength
$\kappa$ and the inverse aspect ratio $\epsilon$ is larger than two.

\section*{Acknowledgments}
The authors gratefully acknowledge discussions with P.~J.~Catto,
P.~Helander, S.~Moradi and Ye.~O.~Kazakov. This work was funded by the
European Communities under Association Contract between EURATOM and
{\em Vetenskapsr{\aa}det}, and by the U.S. DOE under Contract
Nos. DE-FG03-95ER54309 and DE-FG02-07ER54917 as part of the FACETS
SciDAC project and used the resources of the NCCS at ORNL under
Contract No. DEAC05-00OR22725.

\end{document}